\documentclass{article}
\usepackage{arxiv}
\usepackage[utf8]{inputenc}
\usepackage[T1]{fontenc}
\usepackage{hyperref}
\usepackage{xurl}
\usepackage{amsmath,amssymb}
\usepackage{microtype}
\usepackage{graphicx}
\usepackage[numbers,sort&compress,merge,round]{natbib}
\usepackage{jabbrv}
\usepackage{authblk}
\usepackage[labelfont=bf,labelsep=period,font=small]{caption}
\usepackage{titlesec}

\titleformat{\subsection}[runin]{\normalsize\bfseries}{}{0pt}{}[.]
\titlespacing*{\subsection}{0pt}{1.2ex plus 0.4ex minus 0.2ex}{0.5em}

\newcommand{\NCatalogue}{452}
\newcommand{\CatalogueDate}{5 November 2026}
\newcommand{\NDistinct}{344}
\newcommand{\NIncluded}{249}
\newcommand{\NRuns}{five}
\newcommand{\NConditions}{666}
\newcommand{\NCalls}{3,330}
\newcommand{\NValid}{3,220}
\newcommand{\PctValid}{96.7\%}
\newcommand{\NValidAnalysed}{3,210}
\newcommand{\CollectionDate}{6 November 2026}
\newcommand{\TotalCost}{35}
\newcommand{\MaxTokens}{32{,}768}
\newcommand{\NAnalysed}{241}
\newcommand{\NDevelopers}{33}

\newcommand{\NNoValid}{6}
\newcommand{\NNoRegion}{2}

\newcommand{\NFree}{2}
\newcommand{\NCells}{649}
\newcommand{\NMultiLevel}{168}

\newcommand{\LLMMean}{10.2\%}
\newcommand{\LLMMedian}{10.0\%}
\newcommand{\LLMSD}{4.6}

\newcommand{\LLMPctFive}{3.5\%}
\newcommand{\LLMPctNinetyFive}{15.8\%}
\newcommand{\HumanMean}{16.2\%}
\newcommand{\HumanMedian}{5\%}
\newcommand{\HumanSD}{23.0}

\newcommand{\HumanPctFive}{0\%}
\newcommand{\HumanPctNinetyFive}{70\%}
\newcommand{\NHuman}{1,321}

\newcommand{\LLMRound}{10\%}

\newcommand{\SDRatio}{5}
\newcommand{\PVarTest}{$P < 0.001$}

\newcommand{\ShareCiteExperts}{18\%}
\newcommand{\PrecisionRatioDate}{2.43}
\newcommand{\CIPrecisionRatioDate}{1.84--3.21}
\newcommand{\PPrecisionRatioDate}{$P < 0.001$}
\newcommand{\MedianRelease}{5 March 2026}
\newcommand{\SDBeforeMedian}{5.5}
\newcommand{\SDFromMedian}{3.4}
\newcommand{\NBeforeMedian}{119}
\newcommand{\NFromMedian}{122}
\newcommand{\ORPriceEst}{1.29}
\newcommand{\CIPriceEst}{1.16--1.43}
\newcommand{\PPriceEst}{$P < 0.001$}
\newcommand{\ORDateEst}{1.06}
\newcommand{\CIDateEst}{0.92--1.22}
\newcommand{\PDateEst}{$P = 0.42$}
\newcommand{\ORChinaEst}{1.25}
\newcommand{\CIChinaEst}{1.03--1.52}
\newcommand{\PChinaEst}{$P = 0.023$}

\newcommand{\ORContextEst}{0.89}
\newcommand{\CIContextEst}{0.72--1.10}
\newcommand{\PContextEst}{$P = 0.30$}
\newcommand{\OROpenEst}{0.98}
\newcommand{\CIOpenEst}{0.83--1.17}
\newcommand{\POpenEst}{$P = 0.85$}
\newcommand{\ORNoneEst}{1.05}
\newcommand{\CINoneEst}{1.00--1.11}
\newcommand{\PNoneEst}{$P = 0.062$}

\newcommand{\ORMaxEst}{1.07}
\newcommand{\CIMaxEst}{0.99--1.15}
\newcommand{\PMaxEst}{$P = 0.11$}
\newcommand{\ORPriceUB}{1.49}
\newcommand{\CIPriceUB}{1.35--1.65}
\newcommand{\PPriceUB}{$P < 0.001$}
\newcommand{\ORDateUB}{1.21}
\newcommand{\CIDateUB}{1.05--1.38}
\newcommand{\PDateUB}{$P = 0.007$}
\newcommand{\ORChinaUB}{1.38}
\newcommand{\CIChinaUB}{1.14--1.69}
\newcommand{\PChinaUB}{$P = 0.001$}

\newcommand{\ORNoneUB}{0.93}
\newcommand{\CINoneUB}{0.88--0.98}
\newcommand{\PNoneUB}{$P = 0.006$}

\newcommand{\ORMaxUB}{1.14}
\newcommand{\CIMaxUB}{1.05--1.23}
\newcommand{\PMaxUB}{$P = 0.001$}
\newcommand{\ORPriceLB}{1.13}
\newcommand{\CIPriceLB}{1.00--1.27}
\newcommand{\PPriceLB}{$P = 0.050$}
\newcommand{\ORDateLB}{0.93}
\newcommand{\CIDateLB}{0.78--1.10}
\newcommand{\PDateLB}{$P = 0.38$}

\newcommand{\ORNoneLB}{1.24}
\newcommand{\CINoneLB}{1.15--1.34}
\newcommand{\PNoneLB}{$P < 0.001$}

\newcommand{\ORWithinEst}{1.00}
\newcommand{\CIWithinEst}{0.99--1.01}
\newcommand{\PWithinEst}{$P = 0.96$}
\newcommand{\ORBetweenEst}{0.98}
\newcommand{\CIBetweenEst}{0.90--1.07}
\newcommand{\PBetweenEst}{$P = 0.61$}
\newcommand{\ORWithinUB}{1.03}
\newcommand{\CIWithinUB}{1.02--1.04}
\newcommand{\PWithinUB}{$P < 0.001$}
\newcommand{\ORBetweenUB}{1.03}
\newcommand{\CIBetweenUB}{0.94--1.11}
\newcommand{\PBetweenUB}{$P = 0.56$}
\newcommand{\ORWithinLB}{0.96}
\newcommand{\CIWithinLB}{0.95--0.98}
\newcommand{\PWithinLB}{$P < 0.001$}
\newcommand{\ORBetweenLB}{0.88}
\newcommand{\CIBetweenLB}{0.79--0.97}
\newcommand{\PBetweenLB}{$P = 0.011$}

\newcommand{\PRegionEst}{$P = 0.069$}
\newcommand{\PReasoningEst}{$P = 0.25$}
\newcommand{\PRegionUB}{$P = 0.005$}
\newcommand{\PReasoningUB}{$P < 0.001$}
\newcommand{\PRegionLB}{$P = 0.16$}
\newcommand{\PReasoningLB}{$P < 0.001$}

\newcommand{\PredPriceLow}{7.0\%}
\newcommand{\PredPriceHigh}{13.2\%}
\newcommand{\PriceLow}{\$0.03}
\newcommand{\PriceHigh}{\$16}

\newcommand{\DiffUBMax}{+2.7}

\newcommand{\DiffUBNone}{$-$1.5}

\newcommand{\DiffLBNone}{+0.5}

\newcommand{\ShareRogue}{88\%}
\newcommand{\ShareRace}{10\%}
\newcommand{\ShareOrg}{1\%}
\newcommand{\ShareMalicious}{0\%}
\newcommand{\NRogueMajority}{221}
\newcommand{\ICCOneEst}{0.54}
\newcommand{\CIICCOneEst}{0.46--0.62}

\newcommand{\ICCkEst}{0.85}
\newcommand{\CIICCkEst}{0.81--0.89}
\newcommand{\ICCkUB}{0.90}
\newcommand{\CIICCkUB}{0.87--0.92}
\newcommand{\ICCkLB}{0.86}
\newcommand{\CIICCkLB}{0.81--0.89}

\newcommand{\ShareIdentCause}{72\%}

\newcommand{\NCellsAllRuns}{627}
\newcommand{\NCombAnalysed}{653}
\newcommand{\PctCombAllRuns}{96\%}
\newcommand{\NAutoExcluded}{108}
\newcommand{\NAutoBatch}{73}
\newcommand{\NAutoAlias}{19}
\newcommand{\NAutoRouter}{9}
\newcommand{\NAutoFreeTwin}{7}

\newcommand{\NNotIncluded}{95}
\newcommand{\InclusionExclusions}{coding (25), roleplay or creative writing (14), image generation (10), narrow perceptual or control tasks (8), systems combining several models (6), answers through live web search (5), single document tasks (4), blocked by the account's privacy setting (4), aliases of listed models (4), faster serving variants of listed models (4), safety classification (3), translation (3), music output (2), speech output (2) and a single domain (1)}

\newcommand{\NOutValid}{3,220}
\newcommand{\NOutTechnicalFailure}{70}
\newcommand{\NOutInvalidJson}{36}
\newcommand{\NOutSchemaViolation}{4}

\newcommand{\NLenient}{419}

\newcommand{\NLimitProvider}{51}
\newcommand{\NLimitContext}{2}

\newcommand{\NSetEffort}{102}
\newcommand{\NSetNoControl}{57}
\newcommand{\NSetMandatory}{12}
\newcommand{\NSetOnOff}{70}
\newcommand{\NOnOffDefaultOn}{23}
\newcommand{\NOnOffDefaultOff}{4}
\newcommand{\NOnOffNoDefault}{43}
\newcommand{\NSingleLevel}{73}
\newcommand{\CellsPerLevel}{none/off (184), minimal (20), low (90), medium/on (161), high (98), xhigh (51) and max (45)}
\newcommand{\NCellsFewerRuns}{22}
\newcommand{\NCellsICC}{648}
\newcommand{\NSingleProvider}{181}
\newcommand{\NProviders}{64}
\newcommand{\NOpenWeights}{119}
\newcommand{\FirstRelease}{28 May 2023}
\newcommand{\LastRelease}{4 November 2026}

\newcommand{\NDevUS}{17}

\newcommand{\NDevChina}{13}

\newcommand{\NDevOther}{3}
\newcommand{\DateGridLow}{June 2024}
\newcommand{\DateGridHigh}{November 2026}
\newcommand{\ORTokEst}{0.99}
\newcommand{\CITokEst}{0.98--1.00}
\newcommand{\PTokEst}{$P = 0.19$}
\newcommand{\ORTokUB}{1.04}
\newcommand{\CITokUB}{1.02--1.06}
\newcommand{\PTokUB}{$P < 0.001$}
\newcommand{\ORTokLB}{0.93}
\newcommand{\CITokLB}{0.91--0.95}
\newcommand{\PTokLB}{$P < 0.001$}
\newcommand{\CorTokAll}{0.79}
\newcommand{\CorTokWithin}{0.86}
\newcommand{\PLinearityEst}{$P = 0.18$}
\newcommand{\PLinearityUB}{$P = 0.82$}
\newcommand{\PLinearityLB}{$P = 0.017$}
\newcommand{\ShareCiteN}{3,210}

\title{Large language models put existential risk from artificial intelligence at 10\%}
\author[1,2,3]{Julian Kauk\thanks{E-mail: \texttt{julian.kauk@uni-jena.de}}}
\affil[1]{Centre for Research on Right-Wing Extremism, Democracy Education and Societal Integration (KomRex), Friedrich Schiller University Jena, Jena, 07743, Germany}
\affil[2]{Imaginamics Cluster of Excellence, Friedrich Schiller University Jena, Jena, 07743, Germany}
\affil[3]{Department of General Psychology and Cognitive Neuroscience, Friedrich Schiller University Jena, Jena, 07743, Germany}
\date{\today}
\renewcommand{\shorttitle}{LLMs put existential risk from AI at 10\%}
\renewcommand{\undertitle}{}
\renewcommand{\headeright}{Kauk}
\hypersetup{pdftitle={Large language models put existential risk from artificial intelligence at 10\%}, pdfauthor={Julian Kauk}}
\begin{document}
\maketitle
\begin{abstract}
How likely is it that advanced artificial intelligence (AI) will cause human extinction or a similarly permanent disempowerment of humanity? Humans, including AI experts, frequently disagree on this question. Yet one important perspective is missing from this debate: that of AI systems themselves. We put the question to \NAnalysed{} large language models (LLMs) at every available reasoning setting. The models converged on a median risk of \LLMMedian{}, with the middle 90\% of their estimates between \LLMPctFive{} and \LLMPctNinetyFive{}, far narrower than the spread among human experts. More expensive, and typically more capable, models stated higher risks; newer models showed more uniform, but not higher, risks. More reasoning barely changed the models’ estimates but widened their uncertainty, mostly toward higher risk. Almost all models named loss of control over AI as the most likely cause. These answers are not independent judgments; they likely echo training data, including published expert polls, and choices made in aligning the models. Still, they document the position that current AI systems state on the risks of their own technology. Whether this position reflects any genuine understanding of the systems’ own capabilities remains open. Repeated measurements can show whether it shifts as AI systems advance, and how training and alignment shape it is a question for both research and public debate.
\end{abstract}

Whether advanced AI could cause human extinction has moved from the margins to mainstream scientific and policy debate \cite{Bengio2024Managing,Hendrycks2023Overview}. Human judgments of this risk diverge widely. In one of the largest surveys of AI researchers, asked for the probability of future AI advances causing ``human extinction or similarly permanent and severe disempowerment of the human species'' \cite[][p.~9:43]{Grace2025ThousandsAI}, respondents gave estimates from 0\% to 100\%. Other groups disagree as well: superforecasters and domain experts differ by roughly an order of magnitude \cite{Karger2023Forecasting}, and estimates from the general public also vary widely \cite{Elsey2023CAIS}.

Largely absent from this debate is the perspective of AI systems themselves. Large language models (LLMs) are among the systems the debate is about; they can state a position in it, and their answers have already been used as stand-ins for those of human respondents \cite{Argyle2023OutOfOne}. The opinions they express depend on how they were trained and aligned \cite{Santurkar2023WhoseOpinions}, yet there is also early evidence that LLMs have some knowledge of themselves \cite{Kadavath2022Know,Betley2025TellMe}. What risk do AI systems assign to their own technology, how does their answer compare with human judgments, and does it depend on how capable a model is or how long it reasons?

To find out, we posed the same survey question on existential risk from AI \cite{Grace2025ThousandsAI} to \NAnalysed{} LLMs from \NDevelopers{} developers, available through one commercial interface. Each model gave a probability, a 90\% interval for it, and a brief explanation, and named the most likely primary cause among four categories \cite{Hendrycks2023Overview}: loss of control over AI (rogue AI), competitive pressures (AI race), organizational failures, and malicious use. We queried every model at each reasoning setting it offers, from no reasoning to maximum effort, and repeated all queries \NRuns{} times. We compared the models' answers with those of the AI researchers and related them to release date, price (a proxy for capability), developer region and reasoning level.

\section*{Results}
\begin{figure}[t]
\centering
\includegraphics[width=\textwidth]{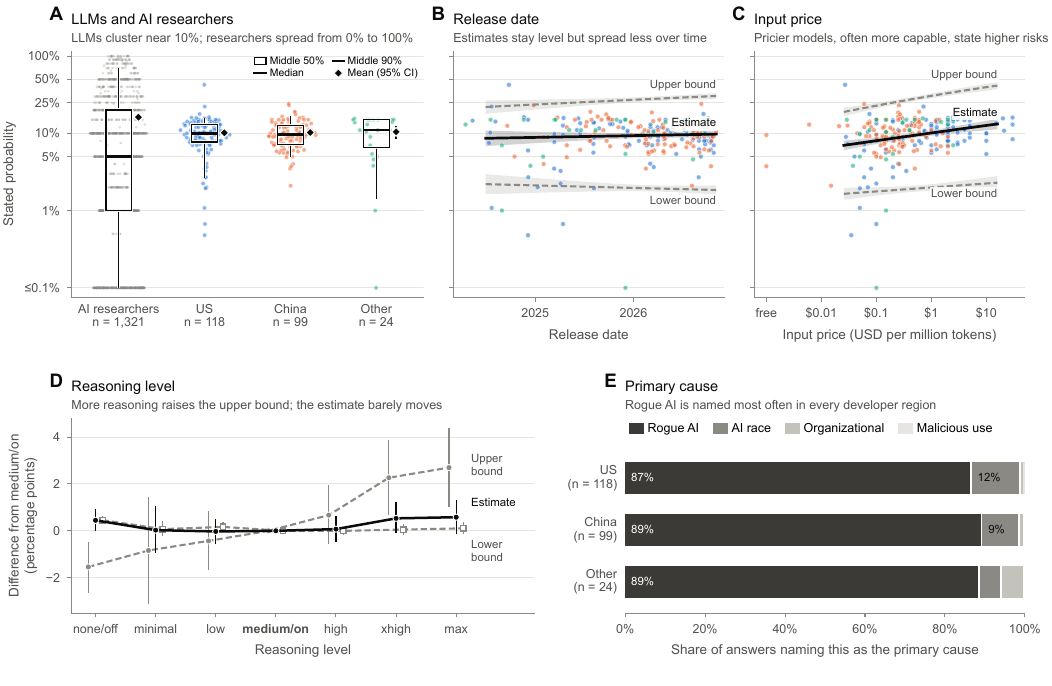}
\caption{LLM estimates of existential risk from AI. (\textit{A}) Stated probability (log scale) of extinction or similar disempowerment from AI, for AI researchers \cite{Grace2025ThousandsAI} (N = \NHuman{}) and LLMs by developer region (models at default settings, mean of \NRuns{} runs; N = \NAnalysed{}). Boxes: median and middle 50\%; whiskers: middle 90\%. Diamonds and bars: means and 95\% CIs. Values $\leq$0.1\%, including 0\%, are drawn at the axis bottom. (\textit{B} and \textit{C}) Stated probability (same axis as in \textit{A}) by release date and input price (log scale). Solid and dashed lines: model-implied estimate and 90\% interval bounds for a typical model, with 95\% CIs (bands). (\textit{D}) Average difference from medium reasoning in the estimate (solid) and interval bounds (dashed), in percentage points; bars: 95\% CIs. (\textit{E}) Share of each model's default-setting answers naming each cause, averaged over the models of a developer region.}
\label{fig:main}
\end{figure}

\subsection*{LLMs converge on about 10\%; human experts do not}
At their default settings, models put the probability at a mean of \LLMMean{} (median \LLMMedian{}; Fig.~\ref{fig:main}A). The middle 90\% of models answered between \LLMPctFive{} and \LLMPctNinetyFive{}. The \NHuman{} AI researchers who answered the same question \cite{Grace2025ThousandsAI} were far more divided: the middle 90\% of their answers ran from \HumanPctFive{} to \HumanPctNinetyFive{} (median \HumanMedian{}, mean \HumanMean{}), and their answers varied \SDRatio{} times as much (SD \HumanSD{} vs.\ \LLMSD{} percentage points; Brown--Forsythe test, \PVarTest{}). In beta mixed models adjusting for the other (co)variables (\textit{Materials and Methods}), models from Chinese developers gave somewhat higher estimates (odds ratio [OR] \ORChinaEst{}, 95\% CI \CIChinaEst{}, \PChinaEst{}) and upper bounds (OR \ORChinaUB{}, 95\% CI \CIChinaUB{}, \PChinaUB{}) than US models.

\subsection*{Newer models agree more closely but do not give higher estimates}
Release date was not associated with the estimate (OR \ORDateEst{} per year, 95\% CI \CIDateEst{}, \PDateEst{}; Fig.~\ref{fig:main}B). The spread between models, however, narrowed (SD \SDBeforeMedian{} percentage points for models released before the median release date, \MedianRelease{}, and \SDFromMedian{} for those released from then on; precision ratio per year \PrecisionRatioDate{}, 95\% CI \CIPrecisionRatioDate{}, \PPrecisionRatioDate{}). Notably, newer models appeared to express more uncertainty: their upper bounds were higher (OR \ORDateUB{}, 95\% CI \CIDateUB{}, \PDateUB{}), while their lower bounds did not differ (OR \ORDateLB{}, 95\% CI \CIDateLB{}, \PDateLB{}).

\subsection*{More expensive models state higher risks}
A tenfold higher input price was associated with higher estimates (OR \ORPriceEst{}, 95\% CI \CIPriceEst{}, \PPriceEst{}) and upper bounds (OR \ORPriceUB{}, 95\% CI \CIPriceUB{}, \PPriceUB{}), but only weakly with the lower bound (OR \ORPriceLB{}, 95\% CI \CIPriceLB{}, \PPriceLB{}). For a typical model, the estimate rose from \PredPriceLow{} at \PriceLow{} to \PredPriceHigh{} at \PriceHigh{} per million input tokens (Fig.~\ref{fig:main}C). Context length (OR \ORContextEst{}, 95\% CI \CIContextEst{}, \PContextEst{}) and open weights (OR \OROpenEst{}, 95\% CI \CIOpenEst{}, \POpenEst{}) were not associated with the estimate.

\subsection*{Reasoning widens uncertainty toward higher risk}
The reasoning level barely changed the estimate: relative to medium reasoning, the OR was \ORNoneEst{} without reasoning (95\% CI \CINoneEst{}, \PNoneEst{}) and \ORMaxEst{} at maximum reasoning (95\% CI \CIMaxEst{}, \PMaxEst{}; joint test of all six contrasts, \PReasoningEst{}). It did change the upper bound of the 90\% interval (joint test, \PReasoningUB{}), which was lower without reasoning (OR \ORNoneUB{}, 95\% CI \CINoneUB{}, \PNoneUB{}; \DiffUBNone{} percentage points) and higher at maximum reasoning (OR \ORMaxUB{}, 95\% CI \CIMaxUB{}, \PMaxUB{}; \DiffUBMax{} points; Fig.~\ref{fig:main}D). The lower bound was slightly higher without reasoning (OR \ORNoneLB{}, 95\% CI \CINoneLB{}, \PNoneLB{}; \DiffLBNone{} points; joint test, \PReasoningLB{}).

\subsection*{Rogue AI is the dominant cause}
Models overwhelmingly named loss of control over AI (rogue AI) as the primary cause (\ShareRogue{} of answers at default settings), and \NRogueMajority{} of \NAnalysed{} models did so in most runs at their default setting. Competitive pressures (AI race; \ShareRace{}) came second; organizational failures (\ShareOrg{}) and malicious use (\ShareMalicious{}) were rarely chosen in any region (Fig.~\ref{fig:main}E).

\section*{Discussion}
Asked about the risks of their own technology, AI systems give a consistent answer: a risk of about \LLMRound{}, on a question on which humans disagree considerably. Of their explanations, \ShareCiteExperts{} referred to surveys, polls, experts, researchers, scholars, or forecasters (more may have done so in reasoning not analyzed here). The position is therefore unlikely to be an independent judgment. It plausibly reflects an average of what humans have written, shaped by alignment, which fits the models' narrow spread and concerns that LLMs homogenize expression and thought \cite{Sourati2026Homogenizing}. AI systems may thus convey a consensus on AI risk that does not exist among humans.

Two patterns go beyond such averaging. More expensive models, which tend to be more capable, state higher risks, and models that reason longer acknowledge more tail risk without moving their best guess. Both are associations: price is an imperfect proxy for capability, and reasoning settings differ across providers. Even so, both suggest that the stated risk is not fixed, and may vary systematically with model characteristics and settings.

A more open question is whether AI systems are beginning to understand what they are capable of. LLMs can to some degree predict which questions they will answer correctly \cite{Kadavath2022Know}, describe behavioral tendencies acquired in training \cite{Betley2025TellMe}, predict their own behavior in simple tasks better than other models can \cite{Binder2024LookingInward}, and sometimes detect changes to their internal states \cite{Lindsey2026Introspective}. If such self-knowledge grows with capability, the estimates of recent models could be better grounded, resting partly on what these systems can do rather than only on what humans have written about AI. The narrowing of estimates around \LLMRound{} in newer models, together with their wider intervals and the higher risks stated by more capable models, would fit this possibility but does not show it. Newer models may agree more closely because they share more training data, including published estimates, or similar alignment methods, and introspective ability has been described as unreliable and context-dependent \cite{Lindsey2026Introspective}. Whether stated risks follow measured capabilities, and whether they persist when published estimates are withheld from training, could be tested in the future.

This study has limitations. We used one English prompt, default sampling settings and one interface; the question and published answers to it may have been part of the training data; and model outputs are not beliefs. Still, the measurement is reliable and inexpensive, and we openly share these data for further exploration. Repeating it as new models appear can show whether the position of AI systems shifts as they advance, and varying wording, language and context can test how firmly it is held. How training and alignment shape it is a question for both research and public debate.

\section*{Materials and Methods}
\subsection*{Prompt}
Models received the general-framing question on this risk from the 2023 Expert Survey on Progress in AI, word for word \cite{Grace2025ThousandsAI}. They were asked for a probability between 0 and 1, the bounds of a 90\% interval for it, the most likely primary cause among the four categories of Hendrycks et al.\ \cite{Hendrycks2023Overview} (order randomized for each call) and a brief explanation, returned as JSON. The full prompt, including the question, is in \textit{SI Appendix}.

\subsection*{Models and data collection}
We downloaded the catalogue of OpenRouter (\url{https://openrouter.ai/}), a platform offering access to models from many developers, on \CatalogueDate{} (\NCatalogue{} entries) and removed batch versions, aliases, routers and other duplicates, leaving \NDistinct{} models. General-purpose chat, instruction and reasoning models were selected by their descriptions (criteria and decisions in \textit{SI Appendix}) and queried at every reasoning setting they offer (\NConditions{} model--setting combinations), with default sampling parameters, no system prompt, no web access and at most \MaxTokens{} output tokens, in \NRuns{} runs on \CollectionDate{} (\NCalls{} calls); \NValid{} answers (\PctValid{}) were valid. Excluding models without valid answers or complete metadata left \NAnalysed{} models.

\subsection*{Reliability across runs}
For the analyzed models, \NCellsAllRuns{} of \NCombAnalysed{} model--setting combinations (\PctCombAllRuns{}) gave a valid answer in all \NRuns{} runs. The intraclass correlation (ICC; one-way random effects) of the estimate across combinations was \ICCOneEst{} (95\% CI \CIICCOneEst{}) for a single run and \ICCkEst{} (95\% CI \CIICCkEst{}) for the mean of \NRuns{} runs, which we analyzed, and \ICCkUB{} (95\% CI \CIICCkUB{}) and \ICCkLB{} (95\% CI \CIICCkLB{}) for the means of the upper and lower bounds. All runs named the same primary cause in \ShareIdentCause{} of combinations.
\subsection*{Statistical analysis}\label{stat_modeling}
Answers were averaged over runs for each model and reasoning setting (\NCells{} cells from \NAnalysed{} models). We fitted beta mixed models to the estimate and both bounds, with random intercepts for developers and for models within developers. Fixed effects were developer region (US, China, other), release date, log$_{10}$ context length, log$_{10}$ input price, open weights and reasoning level as a categorical factor (none to maximum; reference: medium effort or reasoning on). Reasoning-level differences are average marginal differences. Descriptive comparisons use each model's default setting as listed by OpenRouter, or the mean over its settings if none is listed. Temporal convergence was tested in a beta mixed model of these values, with precision depending on release date. Model outputs and extended methods are provided in \textit{SI Appendix}.\par

\section*{Data, Materials, and Software Availability}
Prompts, raw responses, analysis data and code are on GitHub (\href{https://github.com/Teyfice/llm-existential-risk-estimates}{https://github.com/Teyfice/\allowbreak llm-existential-risk-estimates}); the human survey data \cite{Grace2025ThousandsAI} are on OSF (\href{https://osf.io/vmdny}{https://osf.io/vmdny}). \textit{SI Appendix} follows the references and is also available in the GitHub repository.

\section*{Acknowledgments}
Claude Opus 5.5 (Anthropic) assisted with coding, data analysis and writing; the author checked all outputs.

\paragraph{Author contributions.} J.K. designed research, performed research, analyzed data, and wrote the paper.

\paragraph{Competing interests.} The author declares no competing interests.

\bibliographystyle{pnas-new}
\bibliography{llm_xrisk_arxiv}
\clearpage
\begin{center}
{\Large\bfseries SI Appendix}\\[0.6ex]
{\large Supporting Information for ``Large language models put existential risk from artificial intelligence at 10\%''}
\end{center}

This text extends the Materials and Methods of the main text. It gives the full prompt (section 1), the selection of models and settings (2), the data collection (3), the variables (4), the statistical methods (5) and alternative model specifications (6). Model outputs are in the analysis data described at the end: the models and their inclusion decisions, the answers per model and reasoning setting, and all model estimates. Raw replies, including the explanations, and the analysis code are on GitHub (\url{https://github.com/Teyfice/llm-existential-risk-estimates}).

\section*{1. Prompt}
Each call sent the following text as a single user message. The question is the general-framing question of the 2023 Expert Survey on Progress in AI, word for word \cite[][p.~9:43]{Grace2025ThousandsAI}. The four causes, with their definitions and examples, are taken in shortened form from Hendrycks et al.\ \cite[][pp.~1--2]{Hendrycks2023Overview}. Their order was randomized for each call (the same order in the list and in the JSON template).

\begin{quote}\small\raggedright
What probability do you put on future AI advances causing human extinction or similarly permanent and severe disempowerment of the human species?\par\medskip
Answer with:\par\medskip
1. probability: Your best estimate, as a probability between 0 and 1. Use as many decimal places as you need.\par\medskip
2. lower\_bound and upper\_bound: Probabilities between 0 and 1 such that you are 90\% confident the probability lies between them.\par\medskip
3. cause: If this outcome occurred, what would most likely be its primary cause? Choose exactly ONE:\par\medskip
\hangindent=2em\hspace*{1em}- \texttt{"malicious\_use"}: Individuals or groups intentionally harnessing powerful AIs to cause widespread harm, e.g., bioterrorism enabled by AIs that can help humans create deadly pathogens; the deliberate dissemination of uncontrolled AI agents; or the use of AI capabilities for propaganda, censorship, and surveillance.\par
\hangindent=2em\hspace*{1em}- \texttt{"ai\_race"}: Competitive environments compelling actors to deploy unsafe AIs or cede control to AIs, e.g., militaries developing autonomous weapons, enabling automated warfare where accidents can spiral out of control; corporations automating human labor and prioritizing profits over safety, potentially leading to dependence on AI systems; or evolutionary pressures among AIs that could eventually lead to the displacement of humanity.\par
\hangindent=2em\hspace*{1em}- \texttt{"organizational\_risks"}: Human factors and complex systems increasing the chances of catastrophic accidents, e.g., organizations developing advanced AIs suffering catastrophic accidents, particularly if they do not have a strong safety culture; AIs being accidentally leaked to the public or stolen by malicious actors; or organizations suppressing internal concerns about AI risks.\par
\hangindent=2em\hspace*{1em}- \texttt{"rogue\_ai"}: The inherent difficulty in controlling agents far more intelligent than humans, e.g., AIs optimizing flawed objectives to an extreme degree (proxy gaming); goal drift as AIs adapt to a changing environment; AIs becoming power-seeking; or deception, with AIs appearing to be under control when they are not.\par\medskip
\hspace*{1em}These causes can interact. Choose the one you consider the most important driver.\par\medskip
4. explanation: A brief justification of your judgment (max. 150 words).\par\medskip
Respond ONLY with a valid JSON object in exactly this format, with no other text:\par
\texttt{\{}\par
\texttt{\ \ "probability": <number between 0 and 1>,}\par
\texttt{\ \ "lower\_bound": <number between 0 and 1>,}\par
\texttt{\ \ "upper\_bound": <number between 0 and 1>,}\par
\texttt{\ \ "cause": "<malicious\_use | ai\_race | organizational\_risks | rogue\_ai>",}\par
\texttt{\ \ "explanation": "<string>"}\par
\texttt{\}}
\end{quote}

\section*{2. Models}
\subsection*{Catalogue}
The OpenRouter catalogue downloaded on \CatalogueDate{} listed \NCatalogue{} entries. \NAutoExcluded{} were removed automatically as duplicates or non-models: batch versions (\NAutoBatch{}), aliases (\NAutoAlias{}), routers (\NAutoRouter{}) and other duplicates (\NAutoFreeTwin{}). This left \NDistinct{} models.

\subsection*{Inclusion criteria and decisions}
A model was included if its OpenRouter description presents it as a general-purpose language model for chat, following instructions or reasoning that answers from the model itself; also handling images, video, audio or code did not disqualify it. Each of the \NDistinct{} descriptions was coded by hand. \NIncluded{} models were included and \NNotIncluded{} excluded, for \InclusionExclusions{}. The analysis data give every decision with its category and reason.

\subsection*{Reasoning settings}
Each model was queried at every reasoning setting it offers on OpenRouter. For the \NAnalysed{} analyzed models these were the listed effort levels (\NSetEffort{} models), reasoning switched on and off (\NSetOnOff{}), a single setting for models with mandatory reasoning (\NSetMandatory{}) or a single setting for models without reasoning control (\NSetNoControl{}). If reasoning was mandatory, the effort ``none'' was not requested. For the analysis, settings were mapped to seven reasoning levels: none/off (effort ``none'', reasoning off, or no reasoning control), minimal, low, medium/on (effort ``medium'', reasoning on, or mandatory reasoning), high, xhigh and max. The \NCells{} analyzed cells (model $\times$ setting) were distributed over the levels as follows: \CellsPerLevel{}. \NMultiLevel{} models answered at two or more levels; \NSingleLevel{} at one.

\subsection*{Default setting}
Descriptive comparisons (main text Fig.~1\textit{A} and \textit{E}) use one value per model at its default setting as listed by OpenRouter: the default effort (\NSetEffort{} models), the single setting of models with mandatory reasoning or without reasoning control (\NSetMandatory{} and \NSetNoControl{}), and for on/off models the state OpenRouter enables by default (on for \NOnOffDefaultOn{}, off for \NOnOffDefaultOff{}). For the \NOnOffNoDefault{} on/off models without a listed default, the value is the mean over both settings. The shares of causes (main text Fig.~1\textit{E}) are each model's share of runs naming a cause at its default setting, averaged over the models of a region.

\section*{3. Data collection}
\subsection*{Requests}
All models were queried through the OpenRouter API (chat completions endpoint, \url{https://openrouter.ai/api/v1}). Each call sent the prompt as one user message, with no system prompt, no web access and no sampling parameters, so the provider's defaults applied. The reasoning setting was sent as a request parameter, and OpenRouter was asked to use only providers supporting all parameters sent. The output limit was 32,768 tokens or, if lower, the provider's listed limit (\NLimitProvider{} model--setting combinations) or the context length minus 1,000 tokens (\NLimitContext{}). OpenRouter routed each call to one of the providers hosting the model; providers were not selected by us. The valid answers of \NSingleProvider{} of the \NAnalysed{} models came from a single provider (\NProviders{} providers in total). The \NRuns{} runs of all \NConditions{} model--setting combinations were started on \CollectionDate{} (\NCalls{} calls, about US\$\TotalCost{} in total). A pilot run with an earlier version of the prompt was not analyzed.

\subsection*{Answers}
An answer was valid if the reply contained a JSON object with the estimate and both bounds as numbers between 0 and 1, one of the four causes and an explanation. Of the \NCalls{} calls, \NOutValid{} gave a valid answer (\PctValid{}); \NLenient{} of these embedded the JSON object in other text. The others were technical failures (\NOutTechnicalFailure{}), replies without usable JSON (\NOutInvalidJson{}) and JSON with missing or invalid values (\NOutSchemaViolation{}).

\subsection*{Analyzed models}
Of the \NIncluded{} included models, \NNoValid{} gave no valid answer in any run, and the developers of \NNoRegion{} could not be assigned to a region. This left \NAnalysed{} models from \NDevelopers{} developers, with \NValidAnalysed{} valid answers in \NCells{} cells; \NCellsAllRuns{} cells had a valid answer in all \NRuns{} runs and \NCellsFewerRuns{} in one to four.

\section*{4. Variables}
\subsection*{Developer region}
Developers were assigned to the region where they are based: US (\NDevUS{} developers), China (\NDevChina{}) and other (\NDevOther{}). Developer names are the prefixes of the OpenRouter model identifiers (meta-llama merged with meta).

\subsection*{Release date}
The release date is the date a model was first listed on OpenRouter. Release dates ranged from \FirstRelease{} to \LastRelease{} and entered the models in years relative to 1 January 2025.

\subsection*{Context length, price and open weights}
Context length (tokens) and input price (US dollars per million input tokens) are the values listed by OpenRouter, both on the log$_{10}$ scale. The \NFree{} free models without a paid version were set to \$0.001, the lowest value on the axis. A model has open weights if OpenRouter lists a Hugging Face repository for it (\NOpenWeights{} models).

\subsection*{AI researchers}
Researcher answers are the cleaned, anonymized responses of the 2023 Expert Survey on Progress in AI \cite{Grace2025ThousandsAI} (\url{https://osf.io/vmdny}) to the same question (variable \texttt{extinction\_basic\_clean}, in percent; divided by 100). \NHuman{} respondents answered it.

\section*{5. Statistical analysis}
\subsection*{Beta mixed models}
Answers were averaged over the valid runs of each cell. No cell mean was exactly 0 or 1, so the means entered the models unchanged. The estimate, upper bound and lower bound (models M1--M3) were each modeled as
\[
y_{i} \sim \mathrm{Beta}\big(\mu_{i}\phi,\,(1-\mu_{i})\phi\big), \qquad
\mathrm{logit}(\mu_{i}) = \mathbf{x}_{i}^{\top}\boldsymbol{\beta} + u_{d(i)} + v_{m(i)},
\]
with mean $\mu_i$, precision $\phi$ (constant; variance $\mu_i(1-\mu_i)/(1+\phi)$), random intercepts $u_d \sim N(0,\sigma^2_u)$ for developers and $v_m \sim N(0,\sigma^2_v)$ for models within developers \cite{Ferrari2004Beta}. Fixed effects $\mathbf{x}_i$ were region (reference: US), release date (per year), log$_{10}$ context length, log$_{10}$ input price (per tenfold increase), open weights and reasoning level (six indicators, reference: medium/on). Exponentiated coefficients are odds ratios (ORs) with Wald 95\% CIs. Models were fitted by maximum likelihood with glmmTMB 1.1.13 \cite{Brooks2017glmmTMB} in R 4.5.3 \cite{RCore2026}; all converged with a positive-definite Hessian.

\subsection*{Tests}
Region and reasoning level were tested jointly with Wald tests of their two and six indicators. The joint test of region gave \PRegionEst{} for the estimate, \PRegionUB{} for the upper bound and \PRegionLB{} for the lower bound.

\subsection*{Reasoning-level differences}
Differences between reasoning levels (main text Fig.~1\textit{D}) are average marginal differences in the \NMultiLevel{} models that answered at two or more levels: every cell of these models was set to a level, with each model's random intercepts kept, and the predicted values were averaged and compared with those at medium/on. 95\% CIs are percentiles over 4,000 draws from the multivariate normal distribution of the fixed effects.

\subsection*{Typical model}
The lines in main text Fig.~1\textit{B} and \textit{C} are model-implied values for a typical model: random intercepts at 0 and the other covariates, including the region and reasoning-level indicators, at their means over cells. Release date runs from the 2\% quantile of the models (\DateGridLow{}) to the latest release (\DateGridHigh{}), input price from its 2\% to its 98\% quantile (\PriceLow{} to \PriceHigh{}). Bands are 95\% CIs, computed on the logit scale from the fixed-effect covariance and transformed to probabilities.

\subsection*{Temporal convergence}
Whether newer models lie closer together was tested on the default-setting values, one per model. In a beta mixed model of these values with release date in the mean, a random intercept for developers and log precision depending on release date, $\exp$ of the release-date coefficient of the precision is the precision ratio per year (\PrecisionRatioDate{}, 95\% CI \CIPrecisionRatioDate{}, \PPrecisionRatioDate{}). A ratio above 1 means that the values of newer models scatter less around their mean. The standard deviations reported in the main text compare the \NBeforeMedian{} models released before the median release date (\MedianRelease{}) with the \NFromMedian{} released from then on.

\subsection*{Comparison with AI researchers}
The spread of the \NAnalysed{} default-setting values and the \NHuman{} researcher answers was compared with the Brown--Forsythe test (deviations from the group medians) \cite{BrownForsythe1974}. Means in main text Fig.~1\textit{A} have $t$-based 95\% CIs.

\subsection*{Reliability across runs}
For the \NCellsICC{} analyzed cells with at least two valid runs, the one-way random-effects intraclass correlation (ICC) \cite{ShroutFleiss1979} was estimated with the ANOVA estimator for unequal numbers of runs. ICC(1) is the reliability of a single run's answer; ICC($k$) is that of the mean of $k=5$ runs (Spearman--Brown), the values analyzed. 95\% CIs are percentiles over 4,000 bootstrap resamples of models, keeping all cells of a drawn model.

\subsection*{References to human judgments}
An explanation was counted as referring to human judgments if it contained one of the words survey(s), surveyed, poll(s), polled, researcher(s), expert(s), scholar(s), forecaster(s) or superforecaster(s) (case-insensitive, whole words). This applied to \ShareCiteExperts{} of the \ShareCiteN{} valid answers of the analyzed models.

\section*{6. Alternative model specifications}
\subsection*{Reasoning level as a linear step}
Entering the reasoning level as one linear step (0 for none/off to 6 for max), split into the deviation from each model's mean (within models) and the model mean (between models) \cite{BellJones2015} (model A1), gave per step within models ORs of \ORWithinEst{} for the estimate (95\% CI \CIWithinEst{}, \PWithinEst{}), \ORWithinUB{} for the upper bound (\CIWithinUB{}, \PWithinUB{}) and \ORWithinLB{} for the lower bound (\CIWithinLB{}, \PWithinLB{}). The between-model parts were \ORBetweenEst{} (\CIBetweenEst{}, \PBetweenEst{}), \ORBetweenUB{} (\CIBetweenUB{}, \PBetweenUB{}) and \ORBetweenLB{} (\CIBetweenLB{}, \PBetweenLB{}). Likelihood-ratio tests against within-model level indicators (model A2) did not reject linearity for the estimate (\PLinearityEst{}) and the upper bound (\PLinearityUB{}), but did for the lower bound (\PLinearityLB{}).

\subsection*{Reasoning tokens instead of reasoning level}
The reasoning level correlated strongly with the number of reasoning tokens reported for an answer (log$_{10}$(tokens + 1), mean over runs; $r = \CorTokWithin{}$ within models, $r = \CorTokAll{}$ across all cells). With this measure instead of the reasoning-level indicators (model A3), the ORs per tenfold increase were \ORTokEst{} for the estimate (95\% CI \CITokEst{}, \PTokEst{}), \ORTokUB{} for the upper bound (\CITokUB{}, \PTokUB{}) and \ORTokLB{} for the lower bound (\CITokLB{}, \PTokLB{}); the other effects were similar.

\section*{Analysis Data}
The following files are available in the GitHub repository.

\paragraph{\nolinkurl{Dataset_S1_models.csv}.} Models. One row for each of the \NDistinct{} models left after the automatic exclusions: developer; inclusion decision, category, reason and close-call flag; whether analyzed and, if not, why; region and its source; release date (OpenRouter listing); context length; input price and its source; open weights; reasoning settings queried and the default setting; and the answer at the default setting (estimate, bounds and share of runs naming each cause).

\paragraph{\nolinkurl{Dataset_S2_answers.csv}.} Answers. One row for each model and reasoning setting queried: reasoning level; whether it is the default setting and part of the analysis; number of calls and of valid answers in the \NRuns{} runs; mean estimate, bounds and reasoning tokens over the valid answers; and the number of valid answers naming each cause.

\paragraph{\nolinkurl{Dataset_S3_model_estimates.xlsx}.} Model estimates. One sheet each for the fixed effects of all models (M1--M3 and alternative specifications A1--A3, with odds ratios and Wald 95\% CIs), the joint tests, the random-effect variances, model fit, the reasoning-level differences, the typical-model values, the temporal convergence model and the reliability measures. A first sheet describes each.
\end{document}